\documentclass[%
reprint,
amsmath,amssymb, nofootinbib,
aps,
onecolumn
]{revtex4-2}

\usepackage[table,xcdraw]{xcolor}   
\usepackage{amsmath,amssymb,amsthm, graphicx, mathrsfs, pgf,tikz,marvosym,fontawesome,slashed, mathtools,array,bbm,yfonts, adjustbox, adforn, fourier-orns, relsize,stmaryrd,simplewick,youngtab,ytableau,youngtab, epigraph, url, bclogo,arydshln,multirow,dsfont, lipsum, tensor,feynmf, wrapfig, natbib, enumitem,subfig,verbatim}

\usepackage{physics}
\usepackage[left=20mm,right=20mm,top=35mm,bottom=35mm,columnsep=15pt]{geometry} 
\usepackage{placeins,multirow}
\usepackage[T1]{fontenc}
\usepackage[utf8]{inputenc}
\usepackage[nottoc]{tocbibind}
\usetikzlibrary{arrows}
\usetikzlibrary{positioning}
\usepackage[english]{babel}

\usepackage[uppercase, tiny, center]{titlesec}
\usepackage{cancel}
\usepackage{tikz-cd}

\definecolor{r}{cmyk}{1,.50,0,.20} 
\usepackage[colorlinks=true, linkcolor=r, urlcolor=r, linktocpage=true, citecolor=r]{hyperref}

\numberwithin{equation}{section}

\newcommand{\dif}{\textrm{d}}

\setcitestyle{square,comma}
\usepackage{array,mathtools,amssymb,booktabs}
\newcolumntype{C}{>{$}c<{$}}

\begin{document}
	
	\title{\textbf{Gravitational waves in dark bubble cosmology}}
	
	\author{U. Danielsson, D.Panizo}
	\email{ulf.danielsson, daniel.panizo at physics.uu.se}
	\affiliation{Institutionen för Fysik och Astronomi,
		Box 803, SE-751 08 Uppsala, Sweden}
	\author{R.Tielemans}
	\email{rob.tielemans at kuleuven.be}
	\affiliation{Instituut voor Theoretische Fysica, K.U.Leuven,
		Celestijnenlaan 200D, B-3001 Leuven, Belgium
	}

	\begin{flushright}
		UUITP - 03/22
	\end{flushright}

	\begin{abstract}
		In this paper we construct the 5D uplift of 4D gravitational waves in de Sitter cosmology for the brane world scenario based on a nucleated bubble in $AdS_5$. This makes it possible to generalize the connection between the dark bubbles and Vilenkin's quantum cosmology to include gravitational perturbations. We also use the uplift to explain the interpretation of the apparently negative energy contributions in the 4D Einstein equations, which distinguish the dark bubble scenario from Randall-Sundrum.
	\end{abstract}
	
	\maketitle

	\tableofcontents
	
	\newpage

\section{Introduction}\label{sec: Introduction}

Constructing meta-stable de Sitter (dS) vacua in string theory is undoubtedly one of the most challenging problems in contemporary physics that still remains unsolved (see e.g. the arguments in \cite{Danielsson:2018ztv,Obied:2018sgi,vanbeest2021lectures}). In \cite{Banerjee:2018qey} it was proposed that a dS universe can be realized in a string theory setting through a braneworld scenario similar to a Randall-Sundrum (RS) or Karch-Randall (KR) set up \cite{RS1, RS2, KR}. The brane corresponds to an expanding Coleman-de Luccia (CL) bubble of a true vacuum inside a false, unstable, five-dimensional anti de Sitter ($\text{AdS}$) vacuum. The motivation for this model comes from the ubiquity of AdS vacua in string theory which, if non-supersymmetric, are believed to be unstable as a consequence of the weak gravity conjecture \cite{ooguri2017nonsupersymmetric,freivogel2016vacua,2016mtxDanielsson}. The dark bubble scenario thus emerges naturally whenever there is a codimension one brane present in the theory that can mediate the decay of the false $\text{AdS}_5$. Its advantage is twofold: it evades the swampland constraints of constructing `fundamental' dS vacua based on pure compactifications of string theory \cite{Banjeree2021a} and, at the same time, it provides a concrete realisation of a braneworld in string theory with an inside-outside construction (in contrast to the inside-inside structure of RS or KR models). In \cite{Koga:2019yzj,Koga:2020jok,Basile:2021vxh,BasileNoGos} the physics of nucelation in $\text{AdS}_5$ was further studied, while \cite{Berglund2021} discusses a concrete realization of the dark bubble throguh a resolved Calabi-Yau conifold.

The proposed model realises an effective dS vacuum with lower dimensional observers being confined to the bubble boundary, where they perceive an expanding FLRW cosmology with a positive cosmological constant (CC). This model was studied in \cite{Banerjee:2018qey,Banerjee:2019fzz, Banerjee:2020wov, Banerjee:2020wix,Danielsson:2021tyb}. The power of this dark bubble model is that all features in the 4D cosmology will find an interpretation in the bulk:
\begin{itemize}
	\item Four-dimensional gravity arises as an effective description. Indeed, it was shown in \cite{Banerjee:2019fzz} how the 4D Einstein equation follows from the junction condition across the brane. The brane geometry is sourced by the energy-momentum tensor of the brane itself (which for empty branes acts as a cosmological constant) and by contributions from the higher dimensional geometry.
	\item The value of the four-dimensional cosmological constant is set by the differences amongst the AdS scales on the inside and outside of the bubble and by the brane tension. Its positivity is guaranteed by the occurrence of a nucleation event. To construct a phenomenological 4D cosmology with a small vacuum energy, one requires a modest hierarchy where the AdS scales are smaller that the 5D Planck scale.
	\item The familiar dust and radiation components of the 4D cosmic fluid correspond to, respectively, stretched strings and matter in the bulk \cite{Banerjee:2019fzz}.
	\item What a lower-dimensional observer would call `the Big Bang', has the bulk interpretation of a well-understood nucleation event a la Brown-Teitelboim (BT) \cite{Brown:1988kg}. From the higher dimensional perspective, the Big Bang does therefore not appear as a singularity. This was explored in \cite{Danielsson:2021tyb}. 
\end{itemize}

The higher-dimensional interpretation of the Big Bang is particularly interesting as it provides a connection with quantum cosmology. In this inherently 4D framework, one uses canonical quantisation to derive the Wheeler-DeWitt (WdW) equation
\begin{equation}
	\mathcal{H}\Psi[g, \phi] = 0, \label{eq: WdW}
\end{equation}
where $\mathcal{H}$ is the Hamiltonian and $\Psi[g, \phi ] $ is the wave function of the universe defined on the space of all three-geometries $g$, as well as on the space of all matter fields $\phi$ that are present. A common prediction of quantum cosmology is that a spherical universe can spontaneously nucleate out of `nothing', similar to the nucleation event in our dark bubble model. Importantly, any attempt to solve the Schr\"odinger-like equation \eqref{eq: WdW} requires the additional input of boundary conditions, which is a highly non-trivial issue. Two natural choices are Vilenkin's tunneling proposal \cite{VILENKIN198225, VILENKIN1984} and the no-boundary proposal of Hartle and Hawking \cite{HartleHawking}. From the 4D perspective of quantum cosmology neither one seems to be physically favoured\footnote{From a perspective based on Swampland criteria, it is only Vilenkin's proposal that could be realized in string theory \cite{BasileNoGos}.}. In our opinion this should not come as a surprise: boundary conditions should be able to account for the UV behaviour of gravity (for an example see e.g. \cite{Hertog:2021jyd}), but this is not encapsulated by any GR inspired Hamiltonian $\mathcal{H}$. The dark bubble model has the advantage of providing a UV completion of 4D gravity on the bubble making it possible to explore the issue of boundary conditions. Keeping the scale factor as the only dynamical variable, it was shown in \cite{Danielsson:2021tyb} that the BT amplitude in 5D perfectly matches Vilenkin's tunneling amplitude in 4D quantum cosmology.

This tunneling wave function has recently been subject of debate when perturbations about a homogeneous and isotropic background were included. The most natural fluctuations to consider from the point of view of the dark bubble, are of the gravitational type and we refrain from introducing any other kinds of fields. The 4D Hamiltonian is then of the form
\begin{equation}
	\mathcal{H} = \frac{\kappa_4}{24 \pi^2 a}\frac{\partial^2}{\partial a^2} - \frac{\kappa_4}{2a^3}\frac{\partial^2}{\partial h_n^2} - \frac{6 \pi^2}{\kappa_4}\left( a-\frac{\Lambda_4}{3}a^3\right) + \frac{1}{\kappa_4}a (n^2-1)h_n^2,
\end{equation}
where $a$ is the scale factor and $h_n$ represents a specific transverse tracefree tensor mode of $S^3$ labelled by three quantum numbers $(n,\ell, m)$. We have defined $\kappa_4 = 8\pi G_4$. For notational simplicity, mode indices will generically be suppressed. In \cite{Feldbrugge2017, feldbrugge2018inconsistencies, PhysRevD.95.103508,PhysRevD.96.043505,PhysRevLett.121.081302} it was argued that such perturbations are unbounded and would ultimately destroy Vilenkin's quantum cosmology. However, in \cite{Vilenkin2018, Vilenkin:2018oja, DiTucci2019,PhysRevD.100.043544} it was argued that this problem can be avoided by imposing suitable boundary conditions on the perturbation amplitudes. Our embedding of Vilenkin into the dark bubble also suggests that such instability cannot be present since there is no reason why the nucleation event in AdS would be unphysical. 

In this paper we will construct the uplift of 4D gravitational waves to 5D ones. This is necessary if we want to map a mini-superspace model of such perturbations onto a dark bubble nucleation event. For the purpose of writing down the junction condition, one is obliged to find the backreaction of the waves on the bulk geometry. This is a highly non-trivial task that we will tackle perturbatively in the metric
\begin{equation}\label{perturbedMetric}
	g_{\mu\nu} = g_{\mu\nu}^{(0)} + \xi g_{\mu\nu}^{(1)} + \xi^2 g_{\mu\nu}^{(2)} + \mathcal{O}(\xi^3),
\end{equation}
where $\xi$ is a formal expansion parameter. The conventional procedure is that one solves the Einstein equation order by order in $\xi$. Plugging the previous expression into Einstein equation, one finds:
\begin{equation}\label{pertubedEinstein}
	G_{\mu\nu} +\Lambda g_{\mu\nu}= \left(G_{\mu\nu}^{(0)}[g^{(0)}] + \Lambda g_{\mu\nu}^{(0)}\right)+\xi\left(G_{\mu\nu}^{(1)}[g^{(1)}] + \Lambda g_{\mu\nu}^{(1)}\right)+\xi^2\left(G_{\mu\nu}^{(2)}[g^{(1)}]+G_{\mu\nu}^{(1)}[g^{(2)}] + \Lambda g_{\mu\nu}^{(2)}\right) + \mathcal{O}(\xi^3) = 0,
\end{equation}
where $G^{(i)}_{\mu\nu}[g^{(j)}]$ denotes the $i$-th order variation of the Einstein tensor evaluated on the $j$-th order metric perturbation. Formally this is a quantity of order $\max\{i,j\}$ in $\xi$.

At zeroth order, the Einstein equation simply yields the background geometry $g_{\mu\nu}^{(0)}$. In the dark bubble scenario this could, for instance, correspond to a gas of strings inside a Schwarzschild-AdS space.  We will, for simplicity, consider a background of pure AdS.

Gravitational waves (GW) appear at first order in $\xi$ through the linearized Einstein equation (the GW equation), which schematically can be understood as solutions to
\begin{equation}
	G_{\mu\nu}^{(1)}[g^{(1)}] + \Lambda g_{\mu\nu}^{(1)}=0.
\end{equation}
The GW in the dark bubble model must satisfy specific requirements given that the 5D bulk induces a 4D metric on the dark bubble constrained by the junction conditions.

The second order Einstein equation can now be written as
\begin{equation}
	G_{\mu\nu}^{(1)}[g^{(2)}] + \Lambda g_{\mu\nu}^{(2)} = -G_{\mu\nu}^{(2)}[g^{(1)}]. \label{eq: Second order Einstein equation}
\end{equation}
This can in principle be solved to give $g^{(2)}$. This tells us how the geometry reacts to the presence of the gravitational wave $g^{(1)}$. The RHS can be interpreted as an effective energy-momentum tensor $\langle T_{\mu\nu}\rangle\equiv -\kappa_D^{-1}\langle G_{\mu\nu}^{(2)}[g^{(1)}]\rangle$, where $\kappa_D$ is the gravitational constant in $D$ dimensions. This energy-momentum tensor is quadratic in $g^{(1)}$. The angular bracket $\langle\cdot\rangle$ denotes an averaging procedure over several wavelengths that is required for a proper interpretation, see e.g. \cite{Isaacson1968}. Observe the overall minus sign in the definition of this energy-momentum tensor $\langle T_{\mu\nu}\rangle$. We will see the fundamental relevance of this sign's presence through this paper.  This effective energy-momentum term can also be captured by a backreacted background metric. This fact will be of great importance as, once the backreaction is accounted for, the junction condition will dictate how gravitational waves in the bulk will affect the evolution of the four-dimensional bubble.



The outline of the paper is as follows. In section \ref{section: review} we review the dark bubble model and introduce the tools that we need. In section \ref{section: 4Dwaves} we discuss 4D gravitational waves in an expanding universe. In section \ref{section: 5Dwaves} we perform the uplift into the 5D bulk, and perform consistency checks between 4D and 5D. This interplay between the bulk and boundary features will be examined in section \ref{section: Duality}. Finally, we discuss the importance and interpretation of our results.

\section{Review of the dark bubble model}\label{section: review}

\subsection{Friedmann cosmology}

It was proposed in \cite{Banerjee:2018qey} that a $\text{dS}_4$ cosmology can be obtained as the induced 4D metric on a co-dimension one bubble in $\text{AdS}_{5}$. The 5D bulk geometries inside and outside the bubble correspond to $\text{AdS}_{5}$ vacua
\begin{equation}
	\dif s^2_\pm = g^\pm_{\mu\nu}\dif x^\mu\dif x^\nu = -f_\pm(z) \dif t^2_\pm + \frac{\dif z^2}{f_\pm(z)} + z^2\dif \Omega_3^2, \label{eq1: bulk metric}
\end{equation}
where $-(+)$ refers to the inside (outside) of the bubble, $\dif\Omega_3^2=\gamma_{ij}\dif x^i\dif x^j$ is the metric on $S^3$, and $f_\pm$ is for pure $\text{AdS}_5$ given by
\begin{equation}
	f_\pm(z) = 1+k_\pm^2z^2,
\end{equation}
In the following, we will omit the $\pm$ subscript for notational simplicity. The constant $k$ defines the $\text{AdS}_5$ scale (i.e. $L_{\rm AdS} = 1/k$) and the 5D cc is given by $\Lambda_5= -6k^2$. A false (outside) $\text{AdS}_5^+$ vacuum can decay to a true (inside) $\text{AdS}_5^-$ vacuum via the nucleation of a spherical Brown-Teitelboim (BT) instanton \cite{Brown:1988kg} provided $k_->k_+$. Once nucleated, the bubble expands rapidly thereby eating all of $\text{AdS}_5^+$ in a finite time. The bubble can be described by specifying its radius $z=a(\tau)$, where $\tau$ is some time parameter on the bubble. We will assume that the bubble is sufficiently large: $ka\gg1$. The induced metric on the bubble wall is exactly of the FLRW form
\begin{equation}
	\dif s^{2}_{\rm ind} = - N^2(\tau)\dif\tau^{2} + a(\tau)^{2} \dif \Omega_{3}^{2},\label{eq: FLRW}
\end{equation}
where a lapse function $N$ has been introduced to make time reparametrization invariance manifest. The relation between bulk time $t$ and brane time $\tau$ is given by
\begin{equation}
	N^2(\tau) = f(a)\dot{t}^2-\frac{\dot{a}^2}{f(a)} \label{eq: bulk vs brane time}
\end{equation}
where a dot denotes a $\tau$-derivative.

The expansion of the bubble is governed by Israel's junction conditions:
\begin{equation}
	\sigma = \frac{3}{\kappa_{5}}\left(\sqrt{\frac{f_-(a)}{a^2}+\frac{\dot{a}^2}{N^2a^2}}-\sqrt{\frac{f_+(a)}{a^2}+\frac{\dot{a}^2}{N^2a^2}}\right), \label{eq: junction condition}
\end{equation}
where $\sigma$ corresponds to the tension of the bubble wall. By expanding the square root, the first Friedmann equation can be extracted
\begin{equation}
	\frac{1}{N^2}\left(\frac{\dot{a}}{a}\right)^2 = \frac{\kappa_4}{3}\rho_\Lambda - \frac{1}{a^2},
\end{equation}
where the 4D gravitational constant is identified with
\begin{equation}
	\kappa_4=\frac{2k_-k_+}{k_--k_+}\kappa_5,
\end{equation}
and where the 4D cc is determined by $ \rho_\Lambda = \sigma_{\rm cr} -\sigma$ with $\sigma_{\rm cr}$ the critical brane tension at which the bubble remains static
\begin{equation}
	\sigma_{\rm cr} = \frac{3}{\kappa_{5}} (k_{-}-k_{+}).
\end{equation}
The Friedmann equation only admits real solutions if $\sigma<\sigma_{\rm cr}$. From the 5D perspective, this means bubbles with a tension greater than $\sigma_{\rm cr}$ can simply not nucleate.

For a more general bulk metric that corresponds to a gas of strings in a Schwarzschild-AdS space, the function $f$ is given by
\begin{equation}
	f(r) = 1 + k^2r^2 - \frac{\kappa_5M}{3\pi^2 r^2}-\frac{\kappa_5\alpha}{4\pi r}.
\end{equation}
Through the junction condition \eqref{eq: junction condition}, one can identify several different contributions to the Friedmann equation
\begin{equation}
	\frac{1}{N^2}\left(\frac{\dot{a}}{a}\right)^2 = - \frac{1}{a^2} + \frac{\kappa_4}{3}\rho_{\Lambda} + \frac{\kappa_4}{3}\rho_{\rm r}a^{-4}+\frac{\kappa_4}{3}\rho_{\rm m}a^{-3}. \label{eq0: Friedmann}
\end{equation}
The vacuum energy $\rho_\Lambda$, the radiation density $\rho_{\rm r}$ and the matter density $\rho_{\rm m}$ find their origin in the bulk geometry
\begin{align}
	\rho_\Lambda \approx \sigma_{\rm cr}-\sigma&& \rho_{\rm r} \approx \frac{1}{2\pi^2}\left(\frac{M_+}{k_+}-\frac{M_-}{k_-}\right) &&\rho_{\rm m} \approx \frac{3}{8\pi}\left(\frac{\alpha_+}{k_+}-\frac{\alpha_-}{k_-}\right), \label{eq: friedmann components}
\end{align}
We conclude that a bulk black hole with mass $M$ gives rise to radiation in the 4D world, while a gas of stretched strings with average density $\alpha$ gives rise to dust.

\subsection{The general case}

In the following we will use Greek indices when referring to the five-dimensional (bulk) geometry and Latin indices for quantities associated to the induced one. In general, the different bulk metrics across the bubble's wall cause the presence of an energy-momentum tensor $S^{a}_{b}$ on the brane. This can be captured by the (second) Israel's junction condition as:
\begin{equation}
	\kappa_{5}S_{ab} = \left.\left[K_{ab}-Kh_{ab}\right]\right|^-_+, \label{eq: junction condition 2}
\end{equation}
where $[A]_+^- = A_--A_+$, $K_{ab} = \nabla_{\beta} n_{\alpha} \: e^{\alpha}_{a}\: e^{\beta}_{b}$, with $n_{\alpha}$ being an unit normal vector\footnote{Pointing in the direction where the bubble's volume increases.} to the wall and $e_{a}^{\alpha}$ its tangent vector. $h_{ab}$ is the induced metric on the wall. $K_{ab}$ (with trace $K$) represents the extrinsic curvature, which carries information about the bubble's embedding in the bulk geometry. For the sake of simplicity, we will consider the case where the wall is a simple empty brane with $S_{ab}=-\sigma h_{ab}$ with $\sigma$ the brane tension.

To extract the 4D Einstein equations in the general case, one can make use of the Gauss-Codazzi equation
\begin{equation}
	R^{(5)}_{\alpha\beta\gamma\delta}e^\alpha_a e^\beta_b e^\gamma_c e^\delta_d = R^{(4)}_{abcd} + K_{ad}K_{bc}-K_{ac}K_{bd}, \label{GaussCodazzi}
\end{equation}
which connects the extrinsic curvature $K_{ab}$ and the intrinsic curvature of the brane to the projected bulk curvature. Inserting the Gauss-Codazzi equation (and contractions thereof) into the junction condition \eqref{eq: junction condition 2} eliminates the extrinsic curvature in favor of the energy-momentum tensor. Eventually one finds
\begin{equation}
	G_{ab}^{(4)} = \left(\kappa_4\sigma-3k_+k_-\right)h_{ab} + \frac{k_+k_-}{k_--k_+}\left[\frac{\mathcal{J}_{ab}^+}{k_+}-\frac{\mathcal{J}_{ab}^-}{k_-}-\frac{1}{2}\left(\frac{\mathcal{J}^+}{k_+}-\frac{\mathcal{J}^-}{k_-}\right)h_{ab}\right] + \mathcal{O}\left((\kappa_4\Lambda_4)^2\right), \label{ProjectedEinsteinEqs}
\end{equation}
where $\mathcal{J}_{ab}$ is a tensor defined by
\begin{equation}
	\mathcal{J}_{ab} = R^{(5)}_{\alpha\beta\gamma\delta}e^\alpha_a e^\beta_b e^\gamma_c e^\delta_d h^{cd}.
\end{equation}

Thus the four-dimensional geometry is sourced by the bulk geometry through the tensor $\mathcal{J}_{ab}$. One can verify that this expression reproduces the FLRW-case reviewed in the previous section. Note that even in pure $AdS_5$, $\mathcal{J}_{ab}$ has a contribution $-3k^2 h_{ab}$, which contributes to a net cosmological constant given by
\begin{equation}
	\Lambda_4 =6 k_+ k_- - \kappa_4 \sigma = \kappa_4 \left(\frac{3}{\kappa_5}(k_- -k_+ ) - \sigma \right). 
\end{equation}
In the 5D Einstein equation above we see how the bulk geometry induces matter in the effective 4D theory, which then sources the 4D Einstein equations. In \cite{Banerjee:2019fzz} it was shown how localized matter sources in 4D, such as a massive particle, is uplifted into a string that stretches into the bulk, similar to the hanging strings representing quarks in holography. Contrary to Randall-Sundrum \cite{RS1,RS2} or Karch-Randall \cite{KR} models, neither gravity nor matter is localized to the brane but extends holographically into the bulk. It was shown in \cite{Banerjee:2020wov} how the gravitational attraction between two stretched strings in the bulk projects down to the gravitational attraction between two point particles in 4D.

In the rest of the paper we will further study the interplay between 5D and 4D. In particular, we will discuss gravitational waves and their backreaction on the metric. In the dark bubble model the AdS scale $k$ is assumed to be a UV scale that is somewhere between the scales of particle physics and the Planck scale. In the present paper we will for simplicity focus on regimes where the Hubble scale $H$ is much smaller than any such UV-scale, i.e. $H \ll k$. As we will briefly mention, it is in principle possible to relax this assumption.

\section{Gravitational waves in a 4D expanding universe}\label{section: 4Dwaves}

In this section will review gravitational waves in an expanding FLRW cosmology with a flat or spherical topology. When the wavelength is sufficiently small, waves in a flat universe serve as a proxy for those in a spherical universe. Indeed, high frequency waves only probes small regions and do not feel the curvature at larger scales. Gravitational waves are described by transverse-tracefree (TT) perturbations to the metric \eqref{eq: FLRW}. In the conformal time gauge, these are\footnote{Note that in our conventions the coordinates $(\eta,x^i)$ are dimensionless and $a$ has a dimension of length.}
\begin{equation}
	\dif s^2=a^2(\eta)\left[-\dif\eta^2+\left(\gamma_{ij}+\xi h_{ij}(\eta,x)\right)\dif x^i\dif x^j\right], \label{eq: perturbed FLRW}
\end{equation}
where $\gamma_{ij}$ is the metric on a spatial slice in $x^i$-coordinates and $h_{ij}$ is transverse and tracefree. In the following we will, for simplicity, ignore contributions from matter and radiation and consider a pure 4D dS cosmology with positive cosmological constant $\Lambda_4$ only. We then have that the 4D Hubble constant is given by $H^2 = \kappa_4 \rho_\Lambda/3 = \Lambda_4/3$.

\subsection{Flat universe}\label{Flat universe}
The scale factor for a flat dS universe is $a(\eta) = -1/(H\eta)$ with $-\infty<\eta<0$. For concreteness, we will consider a GW travelling in the $x_1$ direction\footnote{It is a trivial modification to the given analysis to consider a wave in any other direction.} with either a $+$ or $\times$ polarization. The perturbation can be expanded into harmonics on the spatial manifold. The first-order Einstein equation then yields a wave equation for each mode separately. For a single mode $h_{\rm 4D}(\eta,x_1) = e^{iqx_1}h_{\rm 4D}(\eta)$, labelled by some continuous wave number $q$, one finds
\begin{equation}
	\frac{\dif^2h_{\rm 4D}}{\dif\eta^2} + 2\mathcal{H}\frac{\dif h_{\rm 4D}}{\dif \eta}+q^2h_{\rm 4D} = 0,
\end{equation}
where $\mathcal{H} = -1/\eta$ is the conformal Hubble rate. Solutions are easily found and given by
\begin{equation}
	h_{\rm 4D}(\eta) = -\eta\cos(q\eta + \phi_0)+\frac{1}{q}\sin(q\eta+\phi_0),
	\label{fourdimensionalsolu}
\end{equation}
where $\phi_0$ is an arbitrary phase. The wave $h$ freezes out to a constant at late times (which happens to be zero if $\phi_0=0$).

\subsection{Closed universe}\label{Closed universe}
A spherical dS universe is a bouncing cosmology with scale factor $a(\eta) = -1/(H\sin\eta)$ with $-\pi/2\leqslant\eta<0$. The moment $\eta= -\pi/2$ is the bounce, which coincides with the moment of nucleation. In a similar vein, one can expand the perturbation in TT harmonics\footnote{We adopt the convention that $S^{3}$ harmonics $Y_{ij}$ satisfy $\triangle Y_{ij} = -(n^2-3)Y_{ij}$ with $n\geqslant3$ with $\triangle$ the Laplacian on $S^3$. Note that mode indices are suppressed. See e.g. \cite{Gerlach, Lindblom}.} on $S^3$. For a single mode $h_{ij}(\eta, x) = h_{\rm 4D}(\eta)Y_{ij}(x)$, labelled by some discrete wave number $n$, the GW equation is
\begin{equation}
	\frac{\dif^2h_{\rm 4D}}{\dif\eta^2} + 2\mathcal{H}\frac{\dif h_{\rm 4D}}{\dif \eta}+(n^2-1)h_{\rm 4D} = 0, \label{eq: Hawking eq}
\end{equation}
where a prime denotes a derivative to conformal time and $\mathcal{H}=a'/a=-\cot\eta$ is the conformal Hubble parameter. It is useful to redefine the time coordinate to
\begin{equation}
	v = \cos\eta,\qquad\qquad \text{with }\:v\:\in\: [0,1). \label{eq: v}
\end{equation}
The GW equation \eqref{eq: Hawking eq} then becomes
\begin{equation}
	\left(1-v^2\right)\frac{\dif^2h_{\rm 4D}}{\dif v^2}+ v\frac{\dif h_{\rm 4D}}{\dif v}+(n^2-1)h_{\rm 4D} = 0.
\end{equation}
With three regular singular points ($v=-1,+1,\infty$), it is well-known that this differential equation can be converted to the hypergeometric kind. The solutions are thus given in terms of these functions:
\begin{subequations}
	\begin{align}
		&h_{\rm 4D}(v)= \:_2F_1\left(-\frac{n+1}{2},\frac{n-1}{2},-\frac{1}{2};1-v^2\right),\\
		&\tilde{h}_{\rm 4D}(v) = \left(1-v^2\right)^{3/2}\:_{2}F_1\left(1-\frac{n}{2},1+\frac{n}{2},\frac{5}{2};1-v^2\right).
	\end{align}%
\end{subequations}
At late times when $v\to 1$, $h$ freezes out to a constant value while $\tilde{h}$ decays completely. Since $n$ is an integer, these hypergeometrics take a simpler form and can be rewritten in terms of the Chebyshev polynomials
\begin{subequations}
	\begin{align}
		&h_{\rm 4D}(v) = vT_n\left(v\right) - \frac{n}{n+1}T_{n+1}\left(v\right),\\
		&\tilde{h}_{\rm 4D}(v) = \sqrt{1-v^2} \left[vU_{n-1}\left(v\right)-\frac{n}{n+1}U_n\left(v\right)\right],
	\end{align}%
	where $T_n$ and $U_n$ are the Chebyshev polynomials of the first and second kind respectively. One can also simplify the Chebyshev polynomials:
	\begin{align}
		&h_{\rm 4D}(v) = \frac{1}{n+1}\cos\left((n+1)\eta\right) +  \sin\eta\sin(n\eta),\\
		&\tilde{h}_{\rm 4D}(v) = \frac{1}{n+1}\sin\left((n+1)\eta\right) - \sin\eta\cos(n\eta),
	\end{align}\label{eq: 4D GW}%
\end{subequations}%
where we used that
\begin{align}
	T_n(v) =\cos(n\eta), && \sqrt{1-v^2}U_n(v) =\sin\left((n+1)\eta\right). \label{eq: Chebyshev}
\end{align}
At late times, high-frequency GWs (large $n$) reduce to \eqref{fourdimensionalsolu} upon the identification $q=\sqrt{n^2-1}\approx n$. This requires taking the limit $\eta\to0$ while keeping $n\eta$ finite. Note that $h_{\rm 4D}$ reduces to a wave with phase $\phi_0=0$ wile $\tilde{h}_{\rm 4D}$ will have the phase $\phi_0=\pi/2$.  Physically, this limit  corresponds to a late-time observer being able to see gravitational fluctuations within his Hubble radius.

\subsection{Energy-momentum tensor}\label{Energy-momentum tensor}

The presence of these GWs affects the background by sourcing an effective energy-momentum tensor as explained in the introduction. Let us first consider this backreaction from a pure 4D perspective. The purpose of this paper is then to explain how this can also be found from the 5D treatment. In particular, GWs in the bulk should backreact on the bubble in the same manner. For phenomenological reasons, we will be interested in the limiting case where the bubble appears to be flat (large wave-number and late time) and where radiation is distributed homogeneously and isotropically throughout the universe. We will therefore restrict this analysis to the waves found in section \ref{Flat universe}.

By following the averaging procedure, i.e. integrating over all the phases $x_1$, one finds the energy-momentum tensor 
\begin{equation}
	\langle \tensor{T}{^a_b}\rangle = \frac{H^2\eta^2}{8\kappa_4}\begin{pmatrix} 7 - 2q^2\eta^2 & \pm2q^2\eta^2 & 0 & 0\\ \mp2q^2\eta^2 & 5+2q^2\eta^2 & 0 & 0\\0&0&1&0\\0&0&0&1 \end{pmatrix},
\end{equation}
where the $\pm$ sign represents waves travelling in opposite $x_1$-directions. An uniform background of gravitational radiation is realised by averaging this tensor over all sorts of waves travelling in all possible $(x_1,x_2,x_3)$-directions with different polarizations. The energy-momentum tensor that describes an uniform background of gravitational radiation is thus given by, in terms of the scale factor,
\begin{equation}
	\langle \tensor{T}{^a_b}\rangle_{\rm iso} = \frac{7}{8\kappa_4}\frac{1}{a^2}\begin{pmatrix}1&0&0&0\\0&\frac{1}{3}&0&0\\0&0&\frac{1}{3}&0\\0&0&0&\frac{1}{3} \end{pmatrix}+\frac{q^2}{4\kappa_4H^2}\frac{1}{a^4}\begin{pmatrix}-1&0&0&0\\0&\frac{1}{3}&0&0\\0&0&\frac{1}{3}&0\\0&0&0&\frac{1}{3}\end{pmatrix}. \label{eq: EM tensor 4D}
\end{equation}
The first term corresponds to a form of energy with equation of state $p=-\rho/3$ whose energy dilutes as $\rho \sim 1/a^2$. This behaves like curvature in the Friedmann equation. The second term has the equation of state $p=\rho/3$ and a dilution $\rho  \sim 1/a^4$, which corresponds to radiation. When the wavelength of gravitational waves is larger than the horizon it becomes frozen, and the curvature component is all that remains.

\subsection{Backreaction}\label{4D backreaction}

To address the problem of backreaction, we will make an Ansatz of what the backreacted geometry should look like. In particular, the geometry will now be sourced by a perturbative amount of radiation and curvature \eqref{eq: EM tensor 4D} without any spatial anisotropies or inhomogeneities. We therefore take the backreacted geometry to be of the form
\begin{equation}
	\dif s^2_{\rm back}= \left(g_{\mu\nu}^{(0)}+g_{\mu\nu}^{(2)}\right)\dif x^\mu\dif x^\nu = a^2(\eta)\left[-\left(1+\xi^2Q(\eta)\right)\dif\eta^2+\gamma_{ij}\dif x^i\dif x^j\right]. \label{eq: backreaction 4D}
\end{equation}
The second order Einstein equation then implies
\begin{equation}
	Q(\eta) = \frac{7}{24}\eta^2 - \frac{1}{12}q^2\eta^4.
\end{equation}
By redefining the time coordinate as
\begin{equation}
	\dif\chi = \sqrt{1+\xi^2 Q(\eta)}\dif\eta \approx \left(1+\frac{1}{2}\xi^2Q(\eta)\right)\dif\eta,
\end{equation}
the metric \eqref{eq: backreaction 4D} is of the FLRW form. Expanding in small $\xi$ one finds
\begin{equation}
	\eta \approx \chi + \xi^2\left(-\frac{7}{144}\chi^3+\frac{1}{120}q^2\chi^5\right).
\end{equation}
By computing the Hubble rate for small $\xi$, one easily recognises contributions from curvature and radiation beyond the dominant cosmological constant
\begin{equation}
	\left(\frac{1}{a^2}\frac{\dif a}{\dif\chi}\right)^2 \approx H^2 + \xi^2\left(-\frac{7H^2}{24}\chi^2+\frac{H^2q^2}{12}\chi^4\right) \approx H^2 + \xi^2\left(-\frac{7}{24}\frac{1}{a^2}+\frac{q^2}{12H^2}\frac{1}{a^4}\right). \label{eq: backreaction 4D Friedmann}
\end{equation}
	
\section{Uplifting gravitational waves to the bulk}\label{section: 5Dwaves}

We are interested in perturbations propagating in the  $\text{AdS}_5$ bulk that correspond to the GW on the brane that were found in the previous section. Clearly, there are different possible fluctuations in the $\text{AdS}_5$ geometry; the ones relevant for the present discussion are TT perturbations to the $S^3$ that enter in the metric as\footnote{Note that in our conventions the coordinates $t$ and $z$ have the dimension of length and the coordinates $x^i$ are --still-- dimensionless.}
\begin{equation}\label{5Dbackgroundplusgw}
	\dif s^2 = -f(z)\dif t^2 + \frac{\dif z^2}{f(z)} + z^2\left(\gamma_{ij}+\xi h_{ij}(t,z,x)\right)\dif x^i\dif x^j,
\end{equation}
where $h_{ij}$ is transverse and tracefree. It is easily checked that the induced metric on the brane in conformal coordinates then corresponds to \eqref{eq: perturbed FLRW}.

\subsection{Finding the 5D wave}\label{Finding the 5D wave}
As before, the TT perturbations can be decomposed in $S^3$ harmonics and for a single mode $h_{ij}=h_{\rm 5D}(\eta,z)Y_{ij}$ one finds the GW equation:
\begin{equation}
	\frac{\partial^2h}{\partial t^2}-f^2\frac{\partial^2h}{\partial z^2}-\frac{f}{z}\left(2+4k^2z^2+f\right)\frac{\partial h}{\partial z} + \frac{n^2-1}{z^2}fh=0, \label{eq: 5D GW equation}
\end{equation}
where we assume emtpy $\text{AdS}_5$ with $f(z)=k^2 z^2 +1$. This determines the evolution of a gravitational wave throughout the $\text{AdS}_5$ bulk. It is useful to work with a coordinate that provides a 5D uplift of \eqref{eq: v}
\begin{equation}
	w = \cos(k t) = \frac{\cos\eta}{\sqrt{1+\left(\frac{H}{k}\right)^2\sin^2\eta}} = \cos\eta + \mathcal{O}\left(\left(\tfrac{H}{k}\right)^2\sin^2\eta\right)
\end{equation}
where the relation between bulk time and conformal time on the brane  \eqref{eq: bulk vs brane time} has been used. This means that in the relevant limit for 4D GR where $H/k\ll1$, one has $w\approx v$. In particular,
\begin{equation}
	kt=\eta + \mathcal{O}\left(\left(\tfrac{H}{k}\right)^2\sin(2\eta)\right)
\end{equation}
Note that this relation is only meaningful once the bubble has nucleated. This occurs at $\eta=-\pi/2$ as alluded to in the 4D treatment. It corresponds to a bulk time $t\approx -\pi/2k$. The bulk time in principle has the full range $-\infty<t<+\infty$. However, one has to take into account the composed inside-outside geometry and the non-eternity of the bubble. The time range of outside geometry is $-\infty<t_+<0$ where the limit $t_+\to0$ corresponds to the bubble having eaten all of $\text{AdS}_5^+$. The inside geometry is only present once a bubble has nucleated after which it persists forever. Therefore $-\pi/2k<t_-<+\infty$.

The GW equation in the bulk is given by
\begin{equation}
	k^2\left[(1-w^2)\frac{\partial^2 h}{\partial w^2}-w\frac{\partial h}{\partial w}\right]-f^2\frac{\partial^2h}{\partial z^2}-\frac{f}{z}\left(2+4k^2z^2+f\right)\frac{\partial h}{\partial z} + \frac{n^2-1}{z^2}fh=0. \label{eq: 5D GW equation}
\end{equation}
This equation needs to be supplemented by suitable boundary conditions. In particular, when $h$ is restricted to the brane (this will be called the `induced wave' $h_{\rm ind}$), we require that $h_{\rm ind}$ coincides with the 4D GW \eqref{eq: 4D GW} found before to leading order in $H/k$. This amounts to imposing the boundary conditions at the location of the bubble $z=a(w)$, assuming $ka\gg1$,
\begin{align}
	h_{\rm ind}(w) \equiv h_{\rm 5D}\left(w,\tfrac{1}{H\sqrt{1-w^2}}\right) = h_{\rm 4D}(v) + \mathcal{O}\left(\left(\tfrac{H}{k}\right)^2\right), && \lim_{z\to0}h_{\rm 5D}(w,z) = 0.
\end{align}
The last condition is the requirement that there are no sources inside the bubble. Note that, in principle, there are two different waves: inside and outside the bubble. It would therefore seem natural to insist that \textit{only} the inside wave decays and \textit{only} the outside wave does not blow up as $z\to\infty$. However note that their evolution is governed by the same wave equation (upto a difference in $k$) and that the boundary condition at the location of the brane must be imposed for both of them. This boundary condition uniquely fixes the inside \textit{and} the outside wave. This means that if the outside wave would be extrapolated in the would-be limit $z\to0$, it would still vanish. 

As far as this uplift is concerned, one may verify that the following meet the requirements
\begin{subequations}
	\begin{align}
		&h_{\rm 5D}(w,z) = \frac{(k z)^{n-1}}{\left(1+k^2z^2\right)^{\frac{n-1}{2}}}\quad\left[wT_n\left(w\right)-\frac{n\left(n+1+2k^2z^2\right)}{2(n+1)\left(1+k^2z^2\right)}T_{n+1}\left(w\right)\right],\\
		&\tilde{h}_{\rm 5D}(w,z) = \frac{(kz)^{n-1}\sqrt{1-w^2}}{\left(1+k^2z^2\right)^{\frac{n-1}{2}}}\left[wU_{n-1}\left(w\right)-\frac{n\left(n+1+2k^2z^2\right)}{2(n+1)\left(1+k^2z^2\right)}U_n\left(w\right)\right].
	\end{align}%
	By using \eqref{eq: Chebyshev}, these uplifted waves can be written as
	\begin{align}
		&h_{\rm 5D}(t,z) = \frac{(k z)^{n-1}}{\left(1+k^2z^2\right)^{\frac{n-1}{2}}}\left[\frac{\frac{1}{2}(1+n)(2-n)+k^2z^2}{(n+1)\left(1+k^2z^2\right)}\cos\left((n+1)kt\right)+\sin(kt)\sin(nkt)\right],\\
		&\tilde{h}_{\rm 5D}(t,z) = \frac{(kz)^{n-1}}{\left(1+k^2z^2\right)^{\frac{n-1}{2}}}\left[\frac{\frac{1}{2}(1+n)(2-n)+k^2z^2}{(n+1)\left(1+k^2z^2\right)}\sin\left((n+1)kt\right)-\sin(kt)\cos(nkt)\right].
	\end{align}
\end{subequations}
Just as for 4D gravitational waves, one might be tempted to take the large $n$, late time ($t\to0$) limit to find an uplift of the GW in a flat universe \eqref{fourdimensionalsolu}. However, one must also take into account that, to reach this flat limit, the wave must be considered near to the bubble. This means it is also required to take the limit in which $kz$ is large; it cannot probe curvature at large scales. The following waves thus represent the correct uplift of \eqref{fourdimensionalsolu}, with phases $\phi_0=0$ and $\phi_0=\pi/2$ respectively, under the identification $q=\sqrt{n^2-1}\approx n$,
\begin{subequations}
	\begin{align}
		&h_{\rm 5D}(t,z) = \frac{-\frac{1}{2}n^2+k^2z^2}{nk^2z^2}\cos\left(nkt\right) + kt\sin(knt),\\
		&\tilde{h}_{\rm 5D}(t,z) = \frac{-\frac{1}{2}n^2+k^2z^2}{nk^2z^2}\sin\left(nkt\right) - kt\cos(knt).
	\end{align}\label{eq: flat 5D waves}%
\end{subequations}%
For large $n$, there are interesting corrections beyond the 4D wave of general relativity at high momentum. At fixed $kz$, there is a competition between $n$ and $kz$ in the first term. The first modification to the 4D wave comes when $kz\sim n$. This translates into $p \sim n/z \sim k$, where $p$ is the proper momentum (or energy) of the wave. We therefore conclude that the AdS-scale $k$ represents a UV-scale where new physics is introduced beyond 4D gravity on the brane.

\subsection{Energy-momentum tensor in the bulk}\label{energy-momentum tensor in the bulk}

In the same spirit as in subsection \ref{Energy-momentum tensor}, we will compute the energy-momentum tensor associated to the waves found before in the flat limit. These waves are described by \eqref{eq: flat 5D waves}. One aims to obtain an isotropic tensor $\langle\tensor{T}{_{\mu}_{\nu}}\rangle_{\rm iso}$ by an average over several wavelengths, polarizations and propagation directions; this superposition of waves represents uniform gravitational radiation filling the bulk geometry. This isotropic stress tensor consists of three identifiable pieces (respectively curvature, radiation, and flux):   
\begin{subequations}
	\begin{equation}
		\langle\tensor{T}{^\mu_\nu}\rangle_{\rm iso} =  \langle\tensor{T}{^\mu_\nu}\rangle_{\rm c}+ \langle\tensor{T}{^\mu_\nu}\rangle_{\rm r}+ \langle\tensor{T}{^\mu_\nu}\rangle_{\rm f}
	\end{equation}
	where each component is given by
	\begin{align}
		&\begin{aligned}\langle\tensor{T}{^\mu_\nu}\rangle_{\rm r} =  \frac{k^2n^2t^2}{4\kappa_5z^2}\begin{pmatrix}-1&0&0&0&0\\0&\frac{1}{3}&0&0&0\\0&0&\frac{1}{3}&0&0\\0&0&0&\frac{1}{3}&0\\0&0&0&0&0 \end{pmatrix}, 
			&&
			\langle\tensor{T}{^\mu_\nu}\rangle_{\rm f} =  \frac{n^2}{8\kappa_5z^2}\begin{pmatrix}0&0&0&0&-\frac{2t}{k^2z^3}\\0&0&0&0&0\\0&0&0&0&0\\0&0&0&0&0\\2k^2tz&0&0&0&0 \end{pmatrix},\end{aligned}\\
		\nonumber\\
		\nonumber&\langle\tensor{T}{^\mu_\nu}\rangle_{\rm c} =   \frac{1}{8\kappa_5z^2}\left(7-\frac{n^4}{2k^4z^4}\right)\begin{psmallmatrix}1&0&0&0&0\\0&\frac{1}{3}+\mathcal{O}\left(\frac{n^2}{k^2z^2}\right)&0&0&0\\0&0&\frac{1}{3}+\mathcal{O}\left(\frac{n^2}{k^2z^2}\right)&0&0\\0&0&0&\frac{1}{3}+\mathcal{O}\left(\frac{n^2}{k^2z^2}\right)&0\\0&0&0&0&1+\mathcal{O}\left(\frac{n^2}{k^2z^2}\right) \end{psmallmatrix}.%
	\end{align}\label{isotropicEnergyStressTensorSecondOrder5D}%
\end{subequations}%
Recalling that the momentum of the wave is $p\sim n/z$, we see how the UV-corrections we commented on earlier, enter in the curvature contribution but not in the radiation piece. The flux contribution shows how these GWs represent a net flow of energy in the positive $z$ direction. We will comment more on this nice feature in section VI.

\subsection{Backreaction}\label{backreaction}

Along similar lines of \ref{4D backreaction}, we will compute the response of the bulk geometry due to the presence of the GWs. We will again limit ourselves to the case where the brane appears flat such that \eqref{eq: flat 5D waves} are the appropriate waves to use. The presence of the energy-momentum tensor \eqref{isotropicEnergyStressTensorSecondOrder5D} will generate a deformation in the bulk's geometry, which can be described by a backreacted metric accounting for it. This backreaction is determined by the second order Einstein equation \eqref{eq: Second order Einstein equation}. To keep the calculation tractable, the GW background was made isotropic in  \eqref{isotropicEnergyStressTensorSecondOrder5D}. To continue approaching this problem from the simplest perspective, it is therefore convenient to adopt the global coordinate system for the backreated bulk geometry. We will make the Ansatz
\begin{equation}\label{backreacted5dmetric}
	\begin{aligned}
		\dif s^2_{\rm back} &= \left(g_{\mu\nu}^{(0)}+ \xi^{2}\:g_{\mu\nu}^{(2)}\right)\dif x^\mu\dif x^\nu  \\
		&\approx -\left[1+k^2z^2+\xi^2\left(q_1-q_2k^2t^2\right)\right]\dif t^2 + \frac{\dif z^2}{1+k^2z^2+\xi^{2}\left(q_1-q_3k^2t^2\right)}+z^2\dif\Omega_3^2.
	\end{aligned}
\end{equation}
The set of coefficients $\{q_i\}$ will be determined later. It is easy to see that in the $\xi\to 0$ limit, we recover the $\text{AdS}_5$ background. The backreaction piece $g_{\mu\nu}^{(2)}$ is given by the $\xi^2$ coefficient in the small $\xi$ expansion, in particular
\begin{align}
	g_{tt}^{(2)} = q_1-q_2k^2t^2, && g_{zz}^{(2)} = \frac{q_3k^2t^2 -q_{1}}{(1+k^2z^2)^2}.
\end{align}
In order to fix the value of the coefficients $\{q_i\}$, one needs to compute the second order Einstein tensor and solve the second order Einstein equation. This yields
\begin{equation}
	q_{1} = -\frac{7}{24},\quad \quad q_{2} = -\frac{q^{2}}{6},\quad \quad q_{3} = \frac{q_{2}}{2} = -\frac{q^{2}}{12}. \label{fixingq}
\end{equation}
This change in the bulk geometry will affect the evolution of the bubble wall at $z=a(\eta)$ through the junction condition \eqref{eq: junction condition}. By computing the extrinsic curvature, one finds (in the large $k$ limit) that the brane's energy-momentum tensor can be written as
\begin{equation}
	\begin{aligned}
		\kappa_4\tensor{S}{^a_b}= -(\kappa_4\: \sigma + \Lambda_{4})\delta^{a}_{b} &+ \delta^{a}_{b}\left(3 H^{2} + \frac{1}{a^{2}}\right) + \frac{2}{a^{2}}\delta^{a}_{0}\delta^{0}_{b} +\\ &+\xi^2 \left[\frac{q_{1}}{a^{2}} (\delta^{a}_{i}\delta^{i}_{b} + 3\: \delta^{a}_{0}\delta^{0}_{b}) + \frac{2q_{2}}{H^{2}a^{4}}\delta^{a}_{i}\delta^{i}_{b} - \frac{3q_{3}}{H^{2}a^{4}}\left(\delta^{a}_{0}\delta^{0}_{b}+\delta^{a}_{i}\delta^{i}_{b}\right)\right],
	\end{aligned}\label{SabProject}
\end{equation}
where $\sigma$ corresponds to the tension of the brane. If we then impose the junction conditions using $S_{ab}=-\sigma h_{ab}$, we obtain the Friedmann equations. Alternatively, one can use the Gauss-Codazzi equation \eqref{GaussCodazzi}, and the projection of Einstein equations \eqref{ProjectedEinsteinEqs} as in \cite{Banerjee:2019fzz}, to obtain the same result in the form
\begin{equation}
	\begin{aligned}
		\tensor{G}{^{(4)}^a_b}=&-\Lambda_4 \delta^{a}_{b} + \xi^2 \left[\frac{q_{1}}{a^{2}} (\delta^{a}_{i}\delta^{i}_{b} + 3\: \delta^{a}_{0}\delta^{0}_{b}) + \frac{2q_{2}}{H^{2}a^{4}}\delta^{a}_{i}\delta^{i}_{b} - \frac{3q_{3}}{H^{2}a^{4}}\left(\delta^{a}_{0}\delta^{0}_{b}+\delta^{a}_{i}\delta^{i}_{b}\right)\right].
	\end{aligned}\label{EnergyMomentumFromProjection}
\end{equation}
Covariant conservation $\nabla_a\tensor{S}{^a_b} = 0$ imposes the same relation between $q_{2}$ and $q_{3}$ that was found in \eqref{fixingq}. This constraint can also be verified by comparing the covariant derivative of the extrinsic curvature\footnote{In the same regime as previous expressions.} to the projection of the bulk energy stress tensor using
\begin{equation}
	\nabla_{a} K_{b}^{a}-\partial_{b} K=G_{\mu \nu} \tensor{e}{^\mu_b} n^{\nu}.
\end{equation}

Note that this result agrees with \eqref{eq: backreaction 4D Friedmann} upon using \eqref{fixingq}. This implies that the junction conditions have taken care of the gravitational perturbation in the bulk, providing a clear connection between the bulk and boundary bubble's cosmological physics.

\begin{figure}[t]
	\centering
	\begin{tikzpicture}[line cap=round,line join=round,>=triangle 45,x=1cm,y=0.75cm]
		\clip(-2,-2.5) rectangle (11,5);
		\draw (0.3,4.9) node[anchor=north west] {$g^{\rm 5D}$};
		\draw [->,-{Computer Modern Rightarrow}, line width=0.3pt] (1.,4.) -- (4.,3.);
		\draw [->,-{Computer Modern Rightarrow},line width=0.3pt] (8.,4.) -- (5.,3.);
		\draw [->,-{Computer Modern Rightarrow},line width=0.3pt,dash pattern=on 1pt off 2pt] (1,4.4) -- (8,4.4);
		\draw [->,-{Computer Modern Rightarrow},line width=0.3pt,dash pattern=on 1pt off 2pt] (1,-0.4) -- (8,-0.4);
		\draw (4.1,5) node[anchor=north west] {\ref{backreaction}};
		\draw (4.1,0.2) node[anchor=north west] {\ref{4D backreaction}};
		\draw (8.2,4.9) node[anchor=north west] {$g^{\rm 5D}_{\rm back}$};
		\draw (4.05,3.3) node[anchor=north west] {$\langle T_{\mu\nu}\rangle$};
		\draw [->,-{Computer Modern Rightarrow},line width=0.3pt] (0.5,4) -- (0.5,0);
		\draw [->,-{Computer Modern Rightarrow},line width=0.3pt] (8.5,4) -- (8.5,0);
		\draw (0.3,0.) node[anchor=north west] {$g^{\rm 4D}$};
		\draw (8.2,0.) node[anchor=north west] {$g^{\rm 4D}_{\rm back}$};
		\draw [->,-{Computer Modern Rightarrow},line width=0.3pt] (1.,-0.8) -- (4.,-1.8);
		\draw [->,-{Computer Modern Rightarrow},line width=0.3pt] (8.,-0.8) -- (5.,-1.8);
		\draw (4.05,-1.5) node[anchor=north west] {$\langle T_{ab}\rangle$};
		\draw (-0.5,2.5) node[anchor=north west] {\ref{Finding the 5D wave}};
		\draw (1.9,-1.2) node[anchor=north west] {\rotatebox{-12}{\ref{Energy-momentum tensor}}};
		\draw (1.9,3.6) node[anchor=north west] {\rotatebox{-12}{\ref{energy-momentum tensor in the bulk}}};
		\draw (6.1,3.6) node[anchor=north west] {\rotatebox{14}{\ref{backreaction}}};
		\path [line width=0.3pt] (9.2,4.4) edge (10,4.4);
		\path [line width=0.3pt] (10,4.4) edge (10,-2);
		\path [line width=0.3pt,-{Computer Modern Rightarrow}] (10,-2) edge (5,-2) ;
		\node at (10,1.5)[right]{\ref{backreaction}};
	\end{tikzpicture}
	\caption{Each section where the computation is performed is indicated by the side of each arrow in the diagram. Horizontal one refers to backreaction, while vertical ones relates the bulk and boundary geometries. Diagonal ones stand for general relativity calculations. The right most line represents the Gauss-Codazzi projection onto the boundary.}
	\label{diagram}
\end{figure}
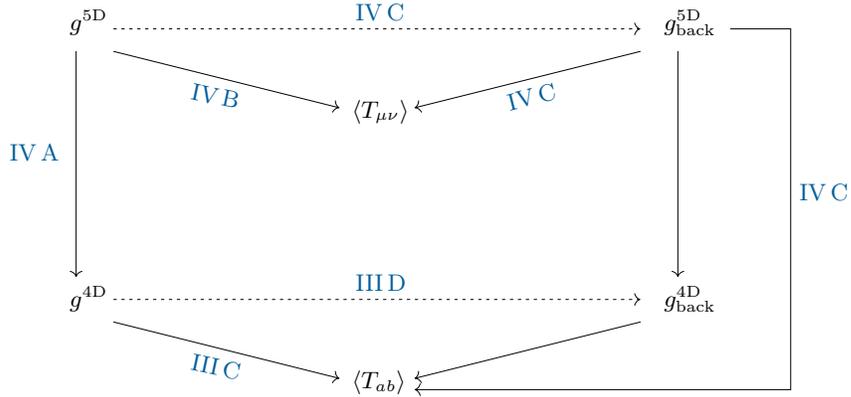

Let us summarize what we have done. It could be useful to navigate through the conceptual diagram in figure \ref{diagram}. In the upper left entry we find the five dimensional background metric plus the gravitational perturbation contribution at first order. The wave extends from $z=0$, where it vanishes, through the interior of the bubble, across the bubble wall and further out through the exterior of the bubble where the wave remains finite. Computing its Einstein equations, averaged over all directions, yields an energy-momentum tensor of the form \ref{isotropicEnergyStressTensorSecondOrder5D}. This same tensor can be obtained if the object sourced by these perturbations is accounted through a "backreacted" bulk geometry at second order as \ref{backreacted5dmetric}.

Restricting $g^{\rm 5D}$ to the boundary of the bubble (i.e $z = \tfrac{-1}{H\eta}$), one can recover the induced four dimensional metric of the expected form $g^{\rm 4D}$, in the conformal time gauge, as shown in  \eqref{eq: perturbed FLRW}. This geometry contains a four dimensional wave that solves the Einstein equations at first order in $\xi$. Solving these at second order, averaging over wavelengths and imposing isotropic superposition, one finds an energy-momentum tensor of the form \eqref{eq: EM tensor 4D}. On the other side of the diagram (upper right corner), starting from the backreacted metric $g^{\rm 5D}_{\rm{back}}$, one can project down this corrected background through the junction conditions to source the Einstein equations in 4D, as shown in \eqref{ProjectedEinsteinEqs}. The tensor $\mathcal{J}_{ab}$ is of the same form on both sides of the brane (up to $k_{\pm}$), but there is a jump in the extrinsic curvature that will backreact on the induced 4D metric. The associated energy-momentum tensor is exactly the same as if you had solved the 4D Einstein equations directly at second order \eqref{eq: EM tensor 4D}, using the averaged 4D waves. This demonstrates the self consistency of the dark bubble model. It is important to realize that it is {\it not} possible to geometrically just project $\langle T_{\mu\nu}\rangle$ from 5D to 4D to obtain $\langle T_{ab}\rangle$. The relation between the two involves the relation between the 5D and 4D gravitational constants, which is determined by the junction conditions.

	\section{The duality between brane dynamics in 5D and Einstein gravity in 4D}\label{section: Duality}
	
	\subsection{Two ways to look at brane dynamics} 
	
	Examining the 4D Einstein equation \eqref{ProjectedEinsteinEqs}, induced by the junction conditions, we note the presence of the tension $\sigma$ and how it acts as a {\it negative} energy density. This is also manifest in \eqref{eq: friedmann components}. This is physically correct: increasing $\sigma$ eventually brings the tension above its critical value yielding a negative 4D cosmological constant that prevents the bubble to nucleate. But what do the fluctuations of the brane correspond to? These would seem to add energy to the tension, thus contributing as a {\it negative} energy density to the 4D energy-momentum tensor. This naively signals an instability. As we will argue, such fluctuations are already taken care of by the junction conditions. 
	
	To see this, one needs to recognise that there are two equivalent ways to describe the motion and fluctuations of the brane. From the 5D perspective one studies the brane equations of motion, as we will see below, which make sure that the backreaction on the 5D geometry is taken into account. It is physically clear that there are no instabilites in this system, beyond the accelerated expansion of the bubble itself, and that all other perturbations will cost energy. On the other hand, one can use the junction conditions \eqref{eq: junction condition 2} as described earlier, where the motion of the brane is captured by 4D Einstein gravity. Eventually, the result should be the same.
	
	If one considers a 5D bulk geometry that induces matter with a positive energy density in 4D, it requires a response from the 4D Einstein tensor with the same positive sign to satisfy the Einstein equations. The brane will give rise to such a geometric contribution that enters into the Einstein equation through the 4D Einstein tensor.  However, one can also view the brane as a contribution to the energy momentum-tensor by formally moving this geometric contribution to the other side of the Einstein equation, thereby picking up a sign. If it is to account for adding matter with a \textit{positive} energy density, the response should be a \textit{negative} energy density. This is precisely what an increase in the brane energy will accomplish. The sign only looks wrong if one, incorrectly, interprets the brane term to be associated with 4D matter. Instead, it is precisely this physical behaviour of the brane that is responsible for 4D gravity. In the case of vibrational modes in the bulk, the brane will start to vibrate and increase its energy as a response. These vibrations are, through the junction conditions, encoded into the 4D Einstein tensor and identical to the response of gravity to a matter source.
	
	The simplest illustration of the two points of view can be found in pure FLRW. We start with the junction  condition \eqref{eq: junction condition} in proper time rewritten as
	\begin{equation} 
		\frac{\sigma}{2} \left(\sqrt{k_{-}^{2}+\frac{\dot{a}^{2}}{a^{2}}+\frac{\Delta_- (a)}{a^2}}+\sqrt{k_{+}^{2}+\frac{\dot{a}^{2}}{a^{2}}+\frac{\Delta_+ (a)}{a^2}}\right) = \frac{3}{2\kappa_{5}}\left(k_{-}^{2}-k_{+}^{2}+\Delta_- (a)-\Delta_+ (a)\right), \label{junction1}
	\end{equation}
	with the metric factor written as $f(r)=k^2 r^2 + \Delta (r)$. Here, we simply have $\Delta_\pm=1$ in pure AdS. Multiplying with the volume of the bubble, proportional to $a^4$, this can be interpreted as energy conservation, comparing the case with and without the bubble. The energy of the brane is simply set equal to the energy difference of the two vacua. Following BT, we note that the energy of the brane itself is given by the average of the energy obtained from the two sides of the brane. This is dictated by the junction conditions. This expression can be viewed as the integrated equations of motion of the brane. If we express the previously mentioned relation using bulk time (taking into account that the time coordinates are different on the two sides), this can be written as
	\begin{equation} 
		\frac{\sigma}{2} \left(  \frac{k_-^2 a^2 + \Delta_-(a)}{a\sqrt{k_-^2 a^2 + \Delta_-(a)-\frac{a'^2}{k_-^2 a^2 + \Delta_-(a)}}} + \frac{k_+^2 a^2 + \Delta_+(a)}{a\sqrt{k_+^2 a^2 + \Delta_+(a)-\frac{a'^2}{k_+^2 a^2 + \Delta_+(a)}}}\right)= \frac{3}{2\kappa_{5}}\left(k_{-}^{2}-k_{+}^{2}+\Delta_- (a)-\Delta_+ (a)\right),\label{junction2}
	\end{equation}
	where the prime denotes a derivative with respect to the bulk time. We recognize the relativistic energy of a brane in a curved background (compare with the energy of a relativistic particle, given by $\frac{m}{\sqrt{1-v^2}}$). Hence, we see how the junction conditions, interpreted as 4D gravity, are equivalent to the equations of motion for the transverse brane degree of freedom, and there are no issues with any wrong sign kinetic terms.

	\subsection{Gravitational fields}
	
	The gravitational waves we have studied in the present paper provide a nice example of the same phenomenon. To see this, consider an oscillator exposed to a gravitational wave that functions as an antenna and absorbs energy. In particular, one can have a situation where the oscillator is in sync with the wave, and sits in an excited state. This is how one should think of the brane in the dark bubble model: in the presence of the 5D gravitational wave the brane becomes excited and this is how the 5D wave backreacts on the brane. Hence, there will be contributions to the energy of the brane corresponding to such excitations that will contribute {\it negatively} in the effective energy density in 4D for the simple reason that they {\it add} to the tension. However, these are precisely the terms needed to cancel the induced energy-momentum tensor in 4D from the 5D waves through the junction conditions \eqref{EnergyMomentumFromProjection}. In fact, this is the way you can argue for that those excitations have to be present.
	
	On the other hand, from our review of the network of backreactions, it is clear how to interpret these terms. If we move them to the other side of Einsteins equations they represent nothing else than the response of 4D gravity to the presence of the waves. The point is that the excitations of the brane can be viewed in two different and dual ways. You either track the detailed and time-dependent fluctuations of the brane, which will carry energy, and show up through the non-trivial Einstein tensor in 4D. This balances the effect of the 5D wave to make sure the junction conditions are solved. Alternatively, you keep the brane in its original position, and let the degrees of freedom of the brane carry the energy.

	\subsection{Gauge fields}
	
	In string theory, this is not the whole story. T-duality requires the existence of further degrees of freedom in the form of gauge fields, which are described by the DBI-action. Their gauge potentials are related to coordinates parallel to the brane, which through T-duality correspond to transverse degrees of freedom of lower dimensional branes. According to T-duality, the number of such degrees of freedom should not change. When excited, they should add energy to the brane tension $\sigma$, and looking at the junction conditions it seems as this should lead to a {\it negative} energy density in 4D. How can this be?
	
	It is crucial to understand that the brane is not a probe. Adding gauge flux to the brane, will force the brane to bend and it will move in response. This will, in turn, affect the bulk geometry, which will give other contributions to the 4D effective energy momentum tensor. As we will argue, the net contribution will always be {\it positive} in physically interesting cases.
	
	An instructive example is the case of charged BPS-strings ending on the brane world. As shown in \cite{1998}, such strings can be described as spikes on the brane world, which act as sources of gauge flux within the brane. The transverse coordinate of the brane, which gives the shape of the spike, also determines the gauge potential and the electric field within the brane. The energy of this configuration is fully captured by the brane action. It is a divergent quantity, but by introducing a cutoff close to the spike one discovers that the energy is proportional to the string tension and the length of the spike up to the cutoff, i.e. the string. This is precisely what one would expect for a BPS-string.
	
	In \cite{1998}, the case of a string going through the brane world was discussed. In particular, what would happen if the string were cut at the brane world, and the end points were to move away from each other? Since the string is BPS one would not expect any net force acting between the end points. The attractive electric force between the endpoints is cancelled by a repulsive force due to the scalar field describing the embedding. Similarly, if you consider two strings on the same side of the brane, pulling upwards, you would expect a repulsive electric force. This is, however, cancelled by an attractive force due to the embedding in analogue with \cite{1998}. In fact, this attractive force is nothing else than the gravitational force as studied in \cite{Banerjee:2020wix} and \cite{Banerjee:2020wov}, where it is interpreted as 4D gravity.
	
	The case of a D3 brane in the background of a stack of D3 branes (which is $\text{AdS}_5 \times S^5$ in the near horizon limit) was discussed in \cite{1999}. The spikes turn into strings with the energy carried by the brane. Similar results were obtained in \cite{2007} for a D5-brane in the same background. Both references considered the brane with the spike being a probe and neglected the backreaction on the bulk. For us the backreaction is crucial. 
	
	By adding a gauge field like in the case above, we get an increased energy density on the brane. One might  worry that this will contribute with the wrong sign in the 4D Einstein equation. However, we must also take into account that the brane deforms into a spike, and that this change in shape backreacts on the bulk geometry. It is now very easy to see what will happen. 
	
\begin{figure}[t]
		\centering
		\includegraphics[width=0.7\textwidth]{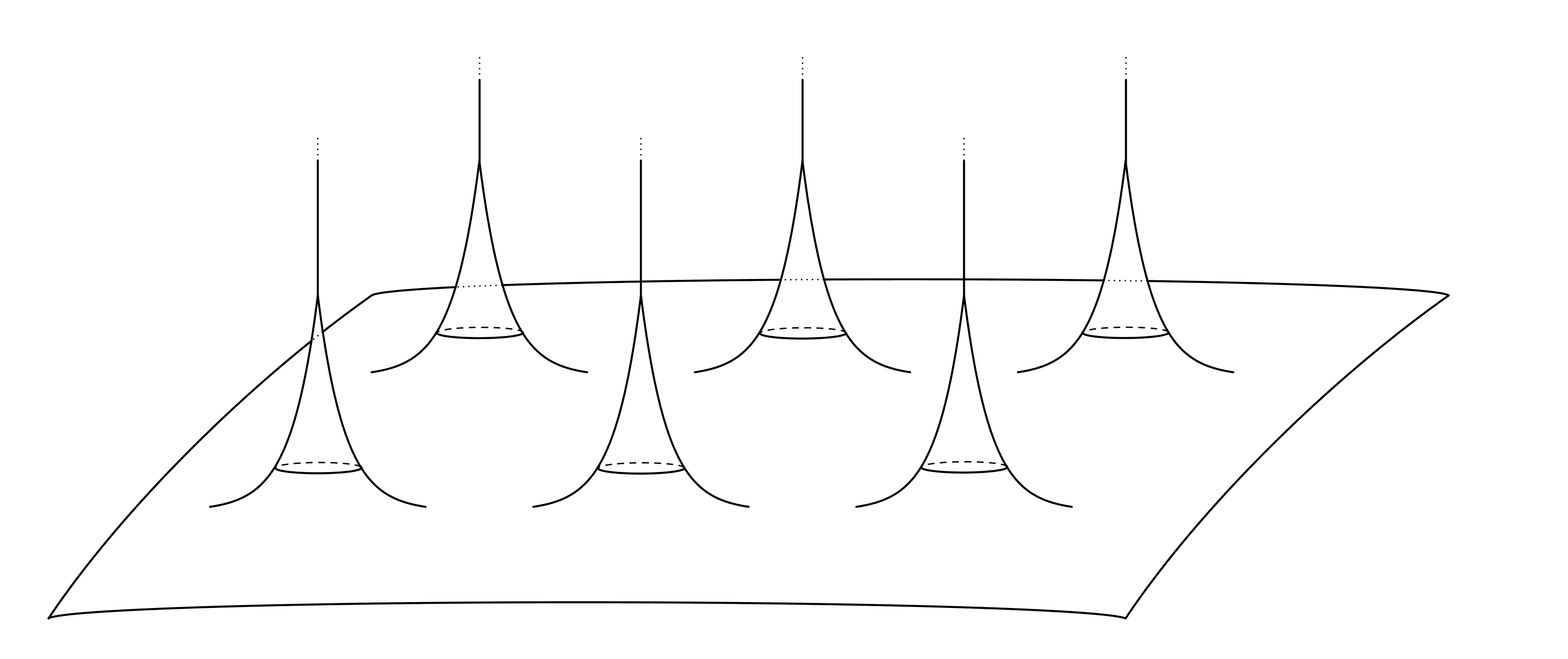}
		\caption{Artistic representation of a distribution of spikes, where the tips of these turns into strings, as if they were pulling upwards.}
		\label{fig: spikes}
	\end{figure}

	For simplicity, let us consider a distribution of such spikes. Relative to the unperturbed piece of the brane in between the spikes, there is a cloud of strings stretching outwards from the bubble. (See figure \ref{fig: spikes}). The gravitational backreaction from the 5D bulk through the junction conditions leads to FLRW with dust, as discussed earlier in the review of the dark bubble model. The net effect of the spikes must therefore be a {\it positive} contribution to the energy density. Hence, our initial conclusion was wrong since we ignored the backreaction.

	Let us summarize what happened. We added matter to the brane, expecting the brane to sag down while yielding a contribution to the energy in 4D with the wrong sign. Instead, the coupling between the gauge field and the scalar forces the brane to bend upwards. This back reacts on the 5D geometry, yielding an extra contribution to the junction condition that pulls the brane up. The net 4D interpretation is a net {\it positive} energy density. 
	
	It would be interesting to study the interplay between the gauge fields on the brane and the physics of the bulk - including 2-form fields to which the strings will couple - to find a 5D uplift of electromagnetic waves and Maxwell-Einstein theory in general. We leave this for future work.

	\section{Discussion and conclusion}\label{Discussion and conclusion}
	
	In this paper we have constructed the uplift of 4D gravitational waves into 5D in the dark bubble model of de Sitter cosmology. The waves extend outwards as well as inwards from the bubble. The waves remains finite everywhere and, in particular, go to zero at the center of the bubble.
	
	The gravitational waves that we have constructed in 5D have an interesting time dependence. Let us, for clarity, focus on waves of high frequency compared to the size of the bubble (this is equivalent to ignoring the positive curvature of the dark bubble universe). In 4D, these reduce to the familiar gravitational waves in a flat universe that at late times exit the horizon and freeze. In 5D, as well as in 4D, there are oscillating terms with coefficients that are constant as well as linear in conformal time $\eta$. The constant piece is what remains after freezing as $\eta \rightarrow 0$. Averaging over many wavelengths, we find constant and quadratic pieces in conformal time in the expressions for the averaged energy density as well as in the back reacted metric. 
	
	Starting out at bulk time $t \sim \eta/k <0$, and increasing $t$, we see that the energy density decreases towards $t=0$, and then starts to increase again when $t$ goes positive. Physically, we have a cloud of radiation that is bound inside the AdS-throat that expands towards maximum dilution and then recollapses. This cloud is matched through the junction conditions to the 4D physics on top of an expanding bubble. Actually, the bubble will expand towards infinity eating up the full AdS as proper time goes to infinity, while conformal time $\eta$ and global bulk time $t$  goes to zero. The recollapsing phase will therefore never occur in the cosmology we study. Clearly, one can envision scenarios with a 4D recollapsing cosmology that would correspond to a recollapsing cloud of radiation in the bulk. Viewed from the bulk, the time scale for the expansion is set by $1/k$ in global time. This is the same as that of oscillating geodesic motion in AdS. Due to the blueshift this translates into cosmological times on the brane world.
	
	The wave that we have constructed is crucial for the applications to quantum cosmology that we initiated in \cite{Danielsson:2021tyb}. There, we studied the WdW equation for a mini-superspace containing only the scale factor, and showed that the dark bubble is a realization of Vilenkin's quantum cosmology. According to our interpretation, it is not quite a creation out of nothing, but a creation out of something, i.e., the already present $\text{AdS}_5$. We argued that our embedding of quantum cosmology into a higher dimension, and the interpretation of the act of creation as simply a CL transition, demonstrates the consistency of the model. The instabilities argued for in \cite{Feldbrugge2017, feldbrugge2018inconsistencies} can therefore not be there on physical grounds.
	
	To make full contact with the Vilenkin version of quantum cosmology, we need to extend the mini-superspace to also include, e.g., gravitational perturbations. Formally, one could do this directly in 4D, trusting that the dark bubble fully reproduce 4D gravity. However, to make use of the higher dimensions to throw new light on the problem, we need the full 5D uplift. This is what we have achieved in the present paper. 
	
	The next step would be to use these waves to investigate the quantum vacuum and its regularization and renormalization starting in 5D. In \cite{Vilenkin:2018oja} a concern in \cite{feldbrugge2018inconsistencies} was addressed, where it was argued that backreaction from the scalars or gravitational modes would destroy the model. As explained in \cite{Vilenkin:2018oja}, this backreaction is nothing else than the vacuum energy of those modes. These contributions need to be taken into account regardless of whether you are studying quantum cosmology or not. Vilenkin introduces a cutoff at fixed proper momentum and obtains a finite result that is absorbed into the cosmological constant.
	
	While this is a standard procedure, which you always need to invoke more or less implicitly when doing cosmology with quantum fields, it is not quite consistent. As reviewed in \cite{2019}, the regularized vacuum energy does not have the correct equation of state, and it is unclear how to treat it. We believe that the uplift to 5D, which is the subject of the present paper, can throw new light on the important problem of quantum contributions to the vacuum energy in an expanding universe. We hope to return to this question in future work.
	
	Finally, we have also commented on the negative sign kinetic terms of brane excitations in the dark bubble model. This sign is a direct consequence of the dark bubble having an inside and an outside, contrary to RS. This changes the physics in a dramatic way. Using a few examples, in particular the gravitational waves, we have shown how the excitations should not be interpreted as 4D matter but instead as the 4D gravitational response mediated by the brane.

	\newpage\newpage
	\begin{acknowledgments}
		We would like to thank Suvendu Giri and Thomas van Riet for comments on an earlier draft. DP would like to thank the Centre for Interdisciplinary Mathematics (CIM) for financial support. The work of RT is supported by the KU Leuven grant C16/16/005 - Horizons in hoge-energie-fysica.
	\end{acknowledgments}
	

	\bibliography{refs}

\end{document}